\documentclass[journal=nale,manuscript=article,10pt]{achemso}
\setkeys{acs}{keywords = true}

\usepackage{chemformula} 
\usepackage[T1]{fontenc} 
\usepackage{graphicx}
\usepackage{pst-solides3d}
\usepackage{tikz,pgfplots}
\usepackage{subcaption}
\pgfplotsset{compat=1.14}
\author{Marco Vettori}
\affiliation{Université de Lyon, Institut des Nanotechnologies de Lyon - INL, UMR CNRS 5270, Ecole Centrale de Lyon, 69134 Ecully, France}
\author{Alexandre Danescu}
\affiliation{Université de Lyon, Institut des Nanotechnologies de Lyon - INL, UMR CNRS 5270, Ecole Centrale de Lyon, 69134 Ecully, France}
\email{alexandre.danescu@ec-lyon.fr}
\author{Xin Guan}
\affiliation{Université de Lyon, Institut des Nanotechnologies de Lyon - INL, UMR CNRS 5270, Ecole Centrale de Lyon, 69134 Ecully, France}
\altaffiliation{Physics Department, Lancaster University, Bailrigg, Lancashire LA1 4YW,UK}
\author{Philippe Regreny}
\affiliation{Université de Lyon, Institut des Nanotechnologies de Lyon - INL, UMR CNRS 5270, Ecole Centrale de Lyon, 69134 Ecully, France}
\author{José Penuelas}
\affiliation{Université de Lyon, Institut des Nanotechnologies de Lyon - INL, UMR CNRS 5270, Ecole Centrale de Lyon, 69134 Ecully, France}
\author{Michel Gendry}
\affiliation{Université de Lyon, Institut des Nanotechnologies de Lyon - INL, UMR CNRS 5270, Ecole Centrale de Lyon, 69134 Ecully, France}

\title[]
{Impact of the Ga flux incidence angle on the growth kinetics of self-assisted GaAs nanowires on Si(111)}

\abbreviations{IR,NMR,UV}
\keywords{GaAs nanowires, molecular beam epitaxy, self-assisted growth, growth modeling, growth kinetics, flux incidence angle, growth modeling}

\begin{document}

%
%
%
%
%

\begin{abstract}
  In this work we show that the incidence angle of group-III elements fluxes plays a significant role on the diffusion-controlled growth of III-V nanowires (NWs) by molecular beam epitaxy (MBE). We present a thorough experimental study on the self-assisted growth of GaAs NWs by using a MBE reactor equipped with two Ga cells located at different incidence angles with respect to the surface normal of the substrate, so as to ascertain the impact of such a parameter on the NW growth kinetics. The as-obtained results show a dramatic influence of the Ga flux incidence angle on the NW length and diameter, as well as on the shape and size of the Ga droplets acting as catalysts. In order to interpret the results we developed a semi-empirical analytic model inspired by those already developed for MBE-grown Au-catalyzed GaAs NWs. Numerical simulations performed with the model allow to reproduce thoroughly the experimental results (in terms of NW length and diameter and of droplet size and wetting angle), putting in evidence that under formally the same experimental conditions the incidence angle of the Ga flux is a key parameter which can drastically affect the growth kinetics of the NWs grown by MBE.
\end{abstract}

\section*{Introduction}
GaAs nanowires (NWs) are one of the most promising materials for the integration of III-V
semiconductors on Si, since they can be grown by molecular beam epitaxy (MBE) on Si substrates
via self-assisted vapor-liquid-solid (VLS) mechanism\cite{Jabeen2008,Cirlin2010,
Krogstrup2010,Plissard_2011,Rudolph2011,Russo-Averchi2015,Dubrovskii2015,Vukajlovic2019} preventing
the use of Au catalyst which would jeopardize the electronic and optoelectronic properties of
these semiconductors, forming deep-level states in both of them\cite{Lang1980,Bar-Sadan2012,
Tambe2010,Breuer2011,Jackson2007,Schmidt2009}. When it comes to MBE, both Au-catalyzed and self-assisted
growths of NWs are diffusion-controlled processes. Many theoretical and experimental studies were carried
out to understand the growth mechanisms and to identify the parameters influencing the NW structure and the growth kinetics\cite{Cirlin2010,Krogstrup2010,Dubrovskii2015,Dubrovskii2011,Yu2012,Glas2013,
Krogstrup2013,Dubrovskii2014,Jacobsson2016,Koivusalo2017,Kim2017,Oehler2018,Harmand2018,
Dubrovskii2005,Dubrovskii2006,Glas2007,Tchernycheva2007,Colombo2008,Dubrovskii2008,Krogstrup2011}.
First works have shown that the catalyst droplet volume or shape\cite{Dubrovskii2011,Yu2012,Krogstrup2013,Glas2007,Krogstrup2011} and, more recently, that the droplet wetting
angle\cite{Jacobsson2016,Kim2017,Harmand2018} do control the NW crystal structure through the
location of the nucleation site. Moreover, the volume of the catalyst droplet controls the
kinetics of the axial growth through the related capture surface\cite{Dubrovskii2005,Dubrovskii2006,Tchernycheva2007}, and
in particular, through the capture area for As in the case of self-assisted GaAs NWs\cite{Krogstrup2013,Kim2017}.

Concerning the NW growth kinetics, the growth models developed so far take into account the Ga flux incidence  angle\cite{Glas2013,Krogstrup2013,Kim2017,Oehler2018,Dubrovskii2005,Dubrovskii2006,Tchernycheva2007,Dubrovskii2008} but do not demonstrate its influence. In particular, the model of Glas {\em et al}\cite{Glas2013} for self-assisted GaAs NWs was based on the assumption that the Ga flux adopted is always high enough to supply the Ga droplet, therefore neglecting to consider the influence of the incidence angle of the Ga flux on the amount of atoms collected by the Ga droplets, and consequently on their volume and shape.

Considering that for self-assisted GaAs NWs: (i) the Ga droplet volume is determined by the balance between the droplet supply in Ga atoms and its depletion caused by the NW growth, and (ii) the droplet supply in Ga occurs through three different ways, i.e. diffusion of Ga adatoms on the substrate, diffusion of Ga adatoms on the NW facets and direct impingement on the droplet, it is expected that the incidence angle of the Ga flux has an influence on the amount of Ga atoms which can be collected by the droplet.

However, despite all these important works, to the best of our knowledge no experimental study has been so far undertaken to ascertain how different Ga flux incidence angles can affect the NW growth kinetics under formally the same growth conditions.

Based on these considerations, we decided to demonstrate experimentally the influence of the Ga flux incidence angle, further denoted $\alpha$, with respect to the surface normal of the substrate (i.e. with respect with the growth axis of vertically grown NWs on the substrate). To this end we used a MBE reactor equipped with two Ga cells at $\alpha \simeq 27.9^\circ$ and $\alpha \simeq 9.3^\circ$ denoted as Ga(5) and Ga(7), respectively. We studied the axial and radial growth rate of GaAs NWs with a series of GaAs NWs grown for different growth times using either the Ga(5) or the Ga(7) cell. The experimental results have been explained by using a semi-empirical model so as to determine the physical factors which originate the significant differences observed with the two different Ga sources.

\section*{Experimental results}

We grew a series of Ga(5)As and Ga(7)As NWs for different growth times ranging from 5 to 80 min, so as to obtain a vast description of the axial growth rate depending on the Ga source used. The growth conditions adopted (cf. - Experimental section) are the same as those one employed in ref. \cite{Vettori2018}, which have proved to provide GaAs NWs with zinc-blende (ZB) structure. SEM images of as-obtained NWs (cf. Figure 1 in Supporting Information (SI)) show that, as expected, despite the equal value of the Ga(5) and Ga(7) fluxes in terms of planar growth rate (0.5 ML/s), the $\alpha$ angle exerts a significant influence on the NW growth kinetics.

Firstly, by observing the SEM images in Figure 1 in Supporting Information (SI), it can be noticed that for shorter growth times (a-f), the lengths of NWs obtained with Ga(5) and Ga(7) cells are comparable, whereas for longer growth times (g-h) the Ga(7)As NWs are significantly shorter than their Ga(5) counterpart. Secondly, a difference in the droplet shape can be observed as the growth time increases (Figure 1). In fact, while for shorter growth times the Ga droplets present equivalent features and wetting angle $\beta$ in the  $138^\circ-142^\circ$ range for both Ga(5) and Ga(7) NWs (Figure 1(a) and 1(b)), for longer growth times the droplets exposed to Ga(7) flux present smaller wetting angle in the $120^\circ-130^\circ$ range (Figure 1(d) and 1(f)) than their Ga(5) counterpart (still in the $138^\circ-142^\circ$ range as shown in Figure 1(c) and 1(e)). Note that the wetting angle $\beta$ is calculated from the relation $R_{NW}=R_d\sin\beta$ with $R_{NW}$ and $R_d$ being the NW and droplet radii, respectively. It can be stated that the droplets on Ga(7)As NWs, contrary to their Ga(5) counterparts, tend to decrease in size as the growth time increases.

The impact of the incidence angle $\alpha$ on the NW growth kinetics is highlighted in Figure 2(a,b) reporting respectively the evolutions of the NW length and NW diameter (measured at the NW top just below the Ga droplet), as a function of the growth time. From Figure 2(a) it can be noticed that while the experimental points for Ga(5)As NWs can be fitted with a single linear regression corresponding to a NW axial growth rate of 1.9 nm/sec, those ones for Ga(7)As NWs lay on the same slope for growth times up to $\simeq 17$ min, but are fitted with a different one for longer growth times, corresponding to a NW axial growth rate of $\simeq 0.8$ nm/sec only. Such a result suggests that in the latter case the growth process undergoes two different growth regimes, named R1 and R2 in Figure 2(a), with a transition from R1 to R2 at a growth time of about 17 min and corresponding to a NW length of about 1.8 $\mu$m (cf. vertical and horizontal dashed blue lines in Figure 2(a)).

\begin{figure}[h!]
  \includegraphics[width=10.5cm]{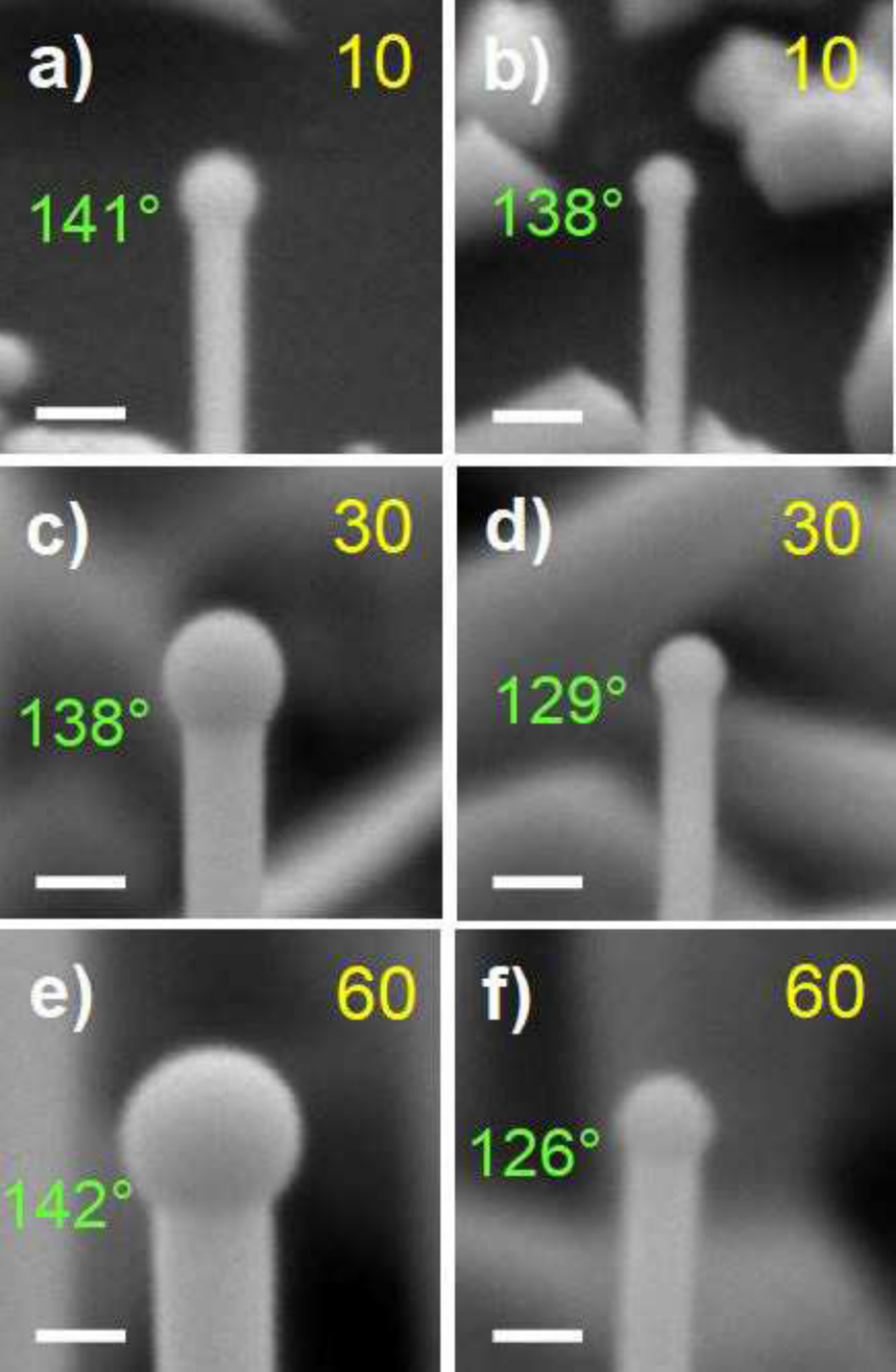}
  \caption{SEM images ($45^\circ$-tilted) showing the evolution of the Ga droplets at the top of Ga(5)As NWs (left) and Ga(7)As NWs (right) with increasing growth time. The growth time in minutes is indicated by yellow numbers. The wetting angle $\beta,$ computed from the identity $R_{NW}=R_d \sin\beta,$ is indicated in green. The white scale bars correspond to 100 nm.}
  \label{fgr:Figure1}
\end{figure} 

\begin{figure}[h!]
  \includegraphics[width=16cm]{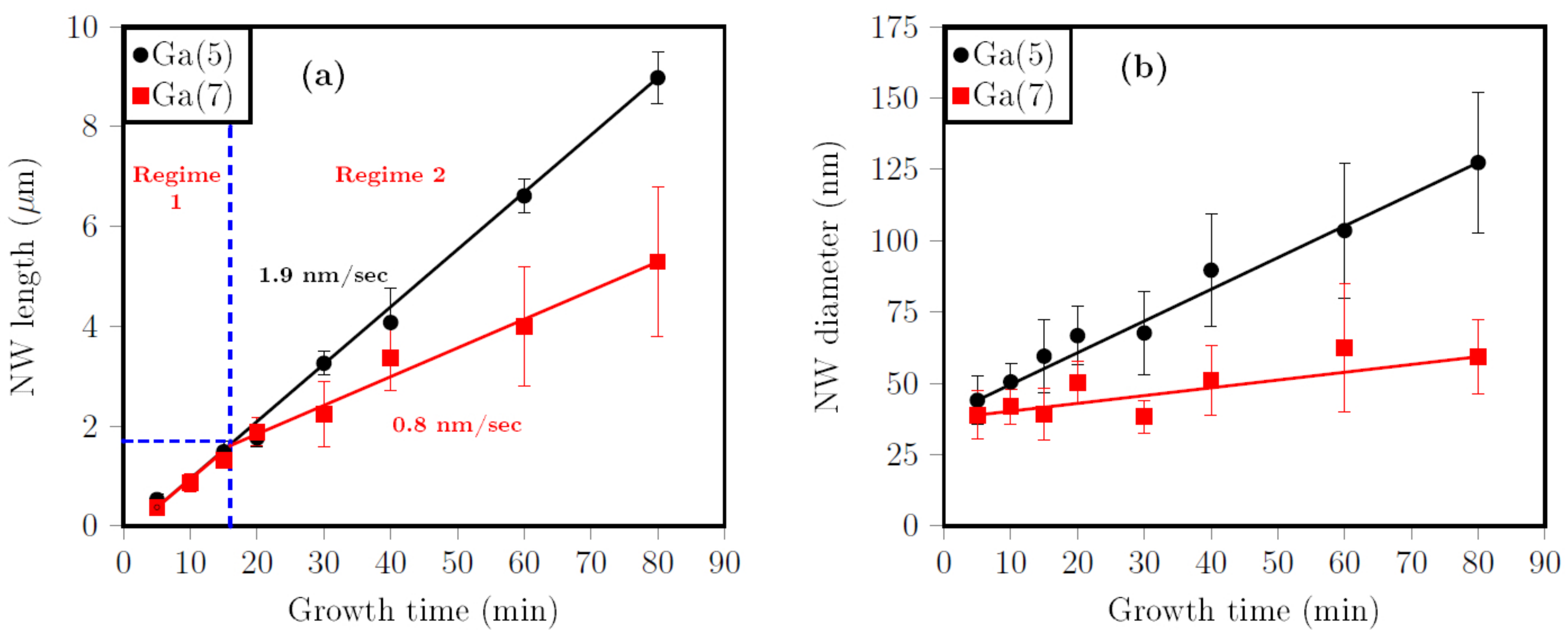}
\caption{(a) Length of Ga(5)As NWs (black) and Ga(7)As NWs (red) as a function of the growth time. The vertical dashed blue line mark the separation between the two growth regimes observed for Ga(7)As NWs, while the horizontal one shows the corresponding NWs length. (b) Diameter of Ga(5)As NWs (black) and Ga(7)As NWs (red) as a function of the growth time.}
  \label{fgr:Figure2}
\end{figure}

As already observable in Figure 1 in SI and clearly highlighted herein in Figure 2(b), $\alpha$ affects not only the NW length evolution with the growth time but also the NW diameter evolution. In particular, while for Ga(5)As NWs the diameter increases linearly with the growth time ($\simeq 1$ nm/min), it seems roughly constant (slope $\simeq 0.2$ nm/min)  in the case of Ga(7)As NWs.

In order to obtain additional insights on the growth process for shorter growth times, a second series of Ga(5)As and Ga(7)As NWs samples was also realized for growth times in the 20 sec - 3 min range (cf. Figure 2 in SI). The results were compared with the first points obtained for longer growth times (Figure 3). As far as the length is concerned (Figure 3(a)), it can be noticed that the linear trend is confirmed for both Ga(5) and Ga(7) cases also at very short growth times, the axial growth rate being still equal to 1.9 nm/sec.

\begin{figure}[h!]
  \includegraphics[width=16cm]{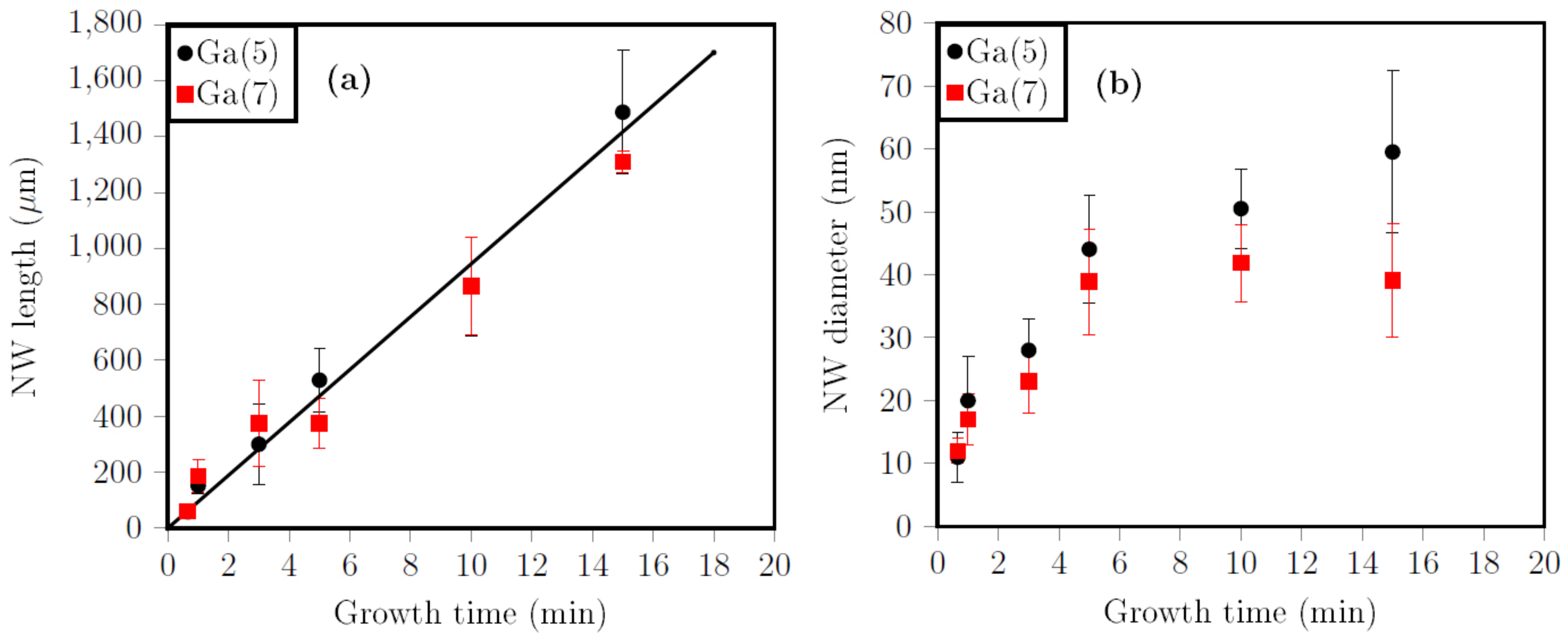}
\caption{Graphics of: (a) the NW length and (b) diameter as a function of the growth time including the data for short growths (20 sec - 3 min). Black and red points correspond to Ga(5)As NWs and Ga(7)As NWs, respectively. The black line in (a) is a guide for the eyes for both black and red points.}
\label{fgr:Figure3}
\end{figure}

On the contrary, Figure 3(b) shows that the trend for the evolution of the diameter is quite different. A rapid increase ($\simeq 5$ nm/min) of the diameter is observed for both Ga(5)As and Ga(7)As NWs for the shortest growth times (20 sec - 5 min), whereas for the longest ones the Ga(5)As NWs show a linear radial growth rate ($\simeq 1$ nm/min), while their Ga(7) counterparts present an almost constant one.  For both cases, the NW diameter increase during the axial growth leads to NWs with an inverse tapered geometry. The different behaviors observed between short and long growth times are thus confirmed by the measure of the tapering coefficient of the NWs T$\%$, as defined by Colombo {\em et al} \cite{Colombo2008}, which results equal to $4-6\%$ with shorter growth times and to $0.5-1\%$ with longer ones, for both Ga(5)As and Ga(7)As NWs. This demonstrates that the radial growth compensating for the tapering effect is higher for the longer growth times (for which the diameter increase is low) than for the short ones (for which such an increase in diameter is higher). It should also be noticed that the NW diameter at the nucleation, occurring after about 12 sec of growth, is $\simeq 15$ nm and corresponds to the average diameter of the Ga droplets as observed before the NW nucleation (cf. Figure 3 in SI).

\section*{Quantitative estimates}

 By extending previous semi-analytical models\cite{Krogstrup2010,Dubrovskii2005,Tchernycheva2007,Dubrovskii2008} proposed for Au-catalyzed and self-assisted III-V NWs, we report a description of the NW growth kinetics using generic Ga and As sources located at $\alpha_Ga$ and $\alpha_As$ angles with respect to the substrate normal. The main original feature of our model is to assume that, in agreement with stability requirements \cite{Gibbs1948}, the wetting angle $\beta$ of the Ga droplet can take only values in an interval $(\beta_{\min},\beta_{\max}),$ and to associate mechanisms to these limit values that allow to the NW diameter below the droplet to increase (or to decrease), resulting in inverse (or direct) tapering.

 In order to identify the model parameters we shall use only the experimental data obtained for the NWs grown by using the Ga(5) source. Then, using these parameters, we simulate the NW growth using the Ga(7) source so as to compare the predicted values for both the axial growth and the changes of the NW diameter (under the droplet) with the above reported experimental data.

\paragraph{The capture surfaces for the Ga atoms:} Based on references \cite{Krogstrup2010,Dubrovskii2005,Tchernycheva2007,Dubrovskii2008} we shall assume three distinct ways for the Ga atoms to supply the droplet (Figure 4): (a) diffusion on the $\textrm{SiO}_2-$terminated Si substrate, (b) diffusion along the NW facets and (c) direct impingement across the droplet surface. These three contributions can be estimated as follows:

\begin{figure}[h!]
  \includegraphics[width=13cm]{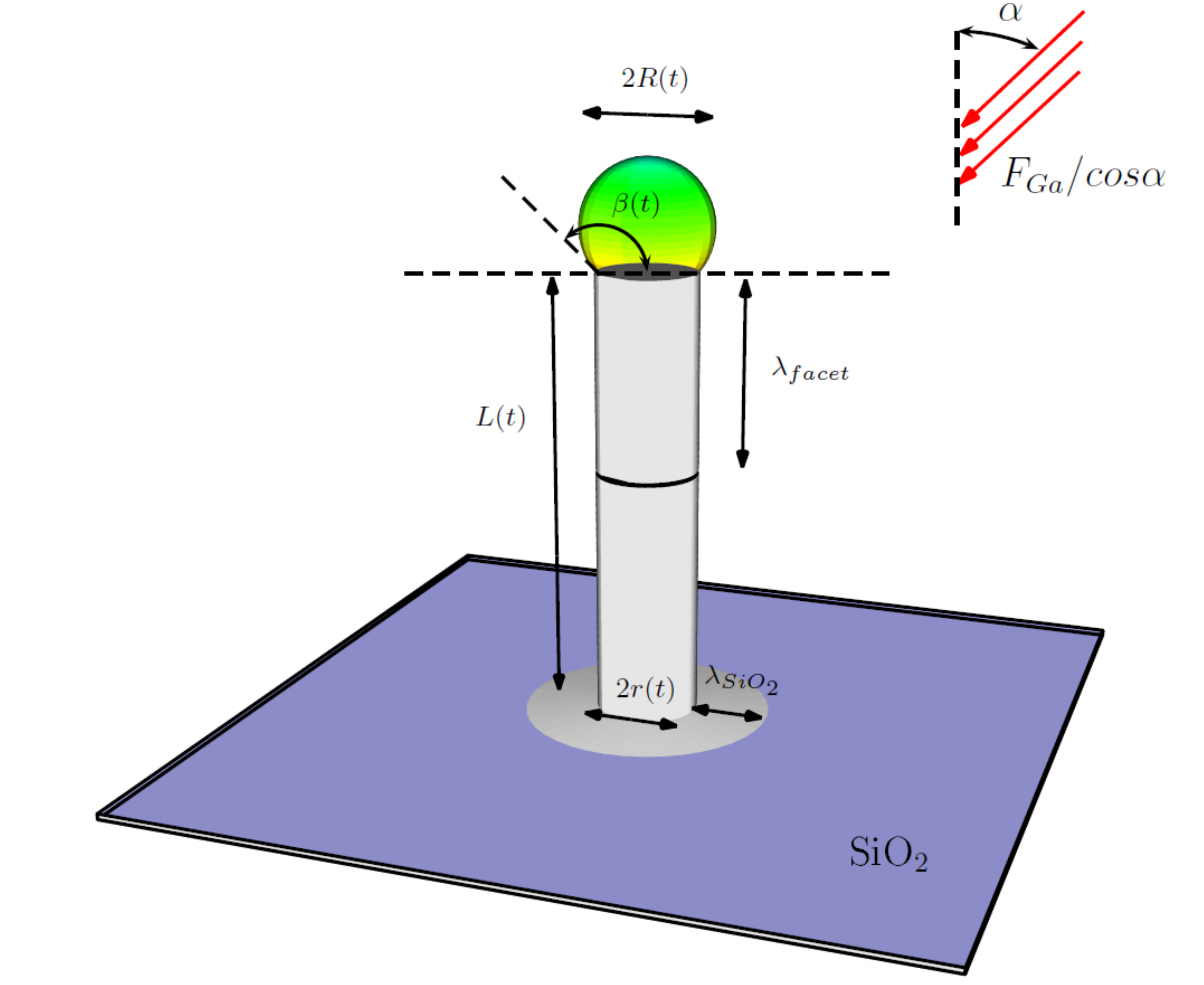}
\caption{Geometry of the NW and droplet, generic flux position at angle $\alpha$ with respect to the substrate normal and diffusion lengths.}
\label{fgr:Figure5}
\end{figure} 

\begin{itemize}
\item[(a)] The amount of Ga atoms, further denoted $q_{Ga}^{sub},$ able to reach the droplet by surface diffusion on the $\textrm{SiO}_2-$terminated Si substrate (which must be followed by diffusion along the NW facets) exists only as long as the NW length, further denoted $L(t),$ is such that $L(t)<\lambda_{\textrm{facet}},$ where $\lambda_{\textrm{facet}}$ corresponds to an {\em average diffusion length on the NW facets}. Per time unit we have
    \begin{equation}
    q_{Ga}^{\textrm{sub}}(t)=F_{Ga}S^{\textrm{sub}}(t),
    \end{equation}
    where $F_{Ga}$ is the Ga flux on the $\textrm{SiO}_2$-terminated Si substrate (fixed at $0.5$ ML/sec), $S^{\textrm{sub}}(t)=\pi[(\lambda_{\textrm{SiO}_2}+r(t))^2-r^2(t)]$ is the substrate capture area, $\lambda_{\textrm{SiO}_2}$ is the average diffusion length on the $\textrm{SiO}_2$-terminated Si substrate and $r(t)$ the NW radius.
\item[(b)] The amount of Ga atoms able to reach the droplet by diffusion along the NW facets can be written as
    \begin{equation}
    q_{Ga}^{\textrm{facet}}(t)= F_{Ga}\tan\alpha_{Ga}S^{\textrm{facet}}(t),\qquad
    \end{equation}
    where $S^{\textrm{facet}}(t) = 2 r(t)min(\lambda_{\textrm{facet}},L(t))$ is the NW facet capture area projected on the plane normal to the Ga flux direction. Here above, the factor $F_{Ga}\tan\alpha_{Ga}$ is the value of the flux on a vertical surface when the nominal flux (i.e. the flux on the plane normal to the direction of the source) is $F_{Ga}/\cos\alpha_{Ga}.$ Finally, the $\min$ function accounts for the NW length $L(t)$ only for NWs with length lower than $\lambda_{\textrm{facet}}.$
\item[(c)] The amount of Ga atoms supplying the droplet by direct impingement is
    \begin{equation}
    q_{Ga}^{\textrm{droplet}}(t)=\frac{F_{Ga}}{\cos\alpha_{Ga}}S(\alpha_{Ga},\beta(t),r(t)).
    \end{equation}
    Here the factor $1/\cos\alpha_{Ga}$ account for the position of the source and the factor $S(\alpha_{Ga},\beta(t),r(t))$ is the exact value of the droplet area projected in the direction normal to the flux when the wetting angle of the droplet is $\beta(t)$ and the droplet is located on top of a NW with radius $r(t),$ as reported by Glas\cite{Glas2010}.
\end{itemize}

\paragraph{The amount of As atoms supplying the droplet:} By using the experimental data for the Ga(5)As NWs and a piecewise linear interpolation for the NW radius and length, we can estimate the amount of As atoms needed to grow the Ga(5)As NWs at $t=80$ min as
$$
\frac{4}{a^3_{GaAs}}\int_{0}^{L_{NW}}\pi r^2(l)\ dl \simeq 1.28\cdot 10^9\ \textrm {atoms},
$$
where $a_{GaAs}$ is the lattice parameter of ZB GaAs. We point out here that the values of the NW diameter reported in Figure \ref{fgr:Figure2} do not include the vapor-solid NW radial growth contribution. Moreover, as the experimental results show that the Ga(5)As NW diameter is constantly increasing, we can deduce that, except at very early stages of the growth process, the droplet wetting angle value equals its maximum one $\beta_{max}$ experimentally measured in the $138^\circ-142^\circ$ range (see Figure 1). With respect to previous models in references \cite{Krogstrup2010,Dubrovskii2005,Tchernycheva2007,Dubrovskii2008}, the existence of a maximum (minimum) value for the droplet wetting angle is a feature of our model that allows including (as described below) a mechanism of increase (decrease) of the NW radius under the droplet.

If the incorporation of As atoms supplying the droplet is the result of only the direct impingement, then knowledge at time $t$ of: the NW radius $r(t)$, the droplet wetting angle $\beta(t)$, the As source incidence angle $\alpha_{As}$ and the nominal $\textrm{As}_4$ flux $F_{As}$ allow a straightforward computation that gives the amount of As atoms, $N_{As},$ supplying the droplet. In our case, an estimation by excess is
$$
N_{As}=F_{As}\int_0^T S(\alpha_{As},\beta_{max},r(t))dt,
$$
where $T$ is the growth-duration and $S(\alpha_{As},\beta_{max},r(t))$ is the projected droplet area on the plane normal to the $\textrm{As}_4$ flux direction, as reported by Glas\cite{Glas2010}. With our numerical data for the Ga(5) source and in agreement with previously reported results in references \cite{Ramdani2012,Glas2013}, we have found that the amount of As atoms supplying the droplet from direct impingement is insufficient for the Ga(5)As NW growth. More exactly, direct impingement provides only $\simeq 89\%$ of the amount of As atoms needed for the NW growth. Thus, we shall follow a previously proposed mechanism\cite{Glas2013} and include also an additional As retro-diffusion flux factor $\epsilon$, so that
\begin{equation}
q_{As}^{droplet}(t)= (1+ \epsilon) F_{As}S(\alpha_{As},\beta(t),r(t)),
\end{equation}
where, from numerical estimates, we take\footnote{The $12.7\%$ missing As atoms are computed with respect to the total amount of As needed; the retro-diffusion coefficient represents the \% of the same quantity with respect to the total amount of As from direct impingement.} $\epsilon=0.127.$

Obviously, the above description of Ga and As sources supplying the droplet holds in an isothermal process at a low density of NWs (in which case the shadowing effects can be neglected).

\paragraph{Growth mechanism description} We shall further assume that there is a critical concentration threshold\cite{Glas2013}, further denoted $c^\star,$ such that solidification occurs only if the droplet concentration $c(t)\geq c^\star$ (over-saturation). The growth process can be described as follows (see the SI):
\begin{enumerate}
\item[1.] At fixed $t$ let $L(t),$ $r(t),$ $\beta(t)$ and $c(t)$ be the NW length, NW radius, droplet wetting angle and As concentration in the droplet, respectively. Size and concentration of the droplet provide the amount of Ga and As atoms in the droplet, further denoted $Q_{Ga}(t)$ and $Q_{As}(t).$ Then, during a small time-interval $(t,t+\Delta t)$ we can update $Q_{Ga}(t)$ and $Q_{As}(t)$ so as to account for the amount of atoms supplying the droplet as described previously:
    \begin{eqnarray}
    Q_{Ga}(t) & \rightarrow & \hat{Q}_{Ga}(t)=Q_{Ga}(t)+(q_{Ga}^{sub}(t)+q_{Ga}^{facet}(t)+q_{Ga}^{droplet}(t))dt,\\
    Q_{As}(t) & \rightarrow & \hat{Q}_{As}(t) = Q_{As}(t) + q_{As}^{droplet}(t)dt.\qquad
    \end{eqnarray}
\item[2.] The knowledge of $\hat{Q}_{Ga}(t)$ and $\hat{Q}_{As}(t)$ provides an estimate for the concentration as:
$$\hat{c}(t) = \hat{Q}_{As}(t)/(\hat{Q}_{Ga}(t)+\hat{Q}_{As}(t))$$
so that, depending on the value of $\hat{c}(t),$ several scenario may occur:
\begin{enumerate}
\item[2.1] The generic case occurs when the updated concentration is such that $\hat{c}>c^\star.$ In this case there is an unique amount $Q(t)=\hat{Q}_{As}(t)\frac{1-c^\star/\hat{c}}{1-2c^\star}$ of equal quantities of Ga and As atoms that can form a crystalline solid phase and such that for the remaining quantities $Q_{Ga}(t+\Delta t)=\hat{Q}_{Ga}(t)-Q(t)$ and $Q_{As}(t+\Delta t)=\hat{Q}_{As}(t)-Q(t)$ we obtain
    \begin{equation}
    c(t+\Delta t)=Q_{As}(t+\Delta t)/(Q_{As}(t+\Delta t)+Q_{Ga}(t+\Delta t))=c^\star.
    \end{equation}
    Thus, both the NW length and diameter do increase with amounts that depend on both the solid material and the remaining liquid quantities: if $Q_{Ga}(t+\Delta t)$ and $Q_{As}(t+\Delta t)$ can form a droplet with $\beta(t+\Delta t)<\beta_{max},$ the NW grows only in the axial direction. If instead $Q_{Ga}(t+\Delta t)$ and $Q_{As}(t+\Delta t)$ cannot form a droplet with radius $r(t)$ and wetting angle $\beta(t+\Delta t)\leq \beta_{max},$ then both the increase of NW radius (under the droplet) and axial growth take place. In this case, the solid phase will modify both the NW radius and the NW length so as to fit the remaining liquid quantities $Q_{Ga}(t+\Delta t)$ and $Q_{As}(t+\Delta t)$ in a droplet with a wetting angle $\beta(t+\Delta t) = \beta_{max}.$
\item[2.2] At the opposite, if $\hat{c} \leq c^\star,$ which may be the case if for instance $q_{Ga}^{sub}(t)+q_{Ga}^{NW}(t)+q_{Ga}^{droplet}(t) > q_{As}^{droplet}(t),$ solidification will not occur but the droplet will change its volume. In this situation, the generic case occurs when the droplet increases its volume at fixed NW radius. But it may happen that $\beta(t)=\beta_{max},$ so that the wetting angle cannot be increased further. In this case, a certain amount of Ga atoms cannot be incorporated into the droplet, because the pinning of the droplet on the NW top is unstable. This situation is very similar to the one encountered when a droplet is supplied by Ga atoms only. In that case, since the wetting angle is bounded by $\beta_{max},$ incorporation of Ga atoms in the droplet stops at this value of the wetting angle. Decreasing the amount of Ga atoms that can be incorporated in the time-interval $(t,t+\Delta t)$ increases the concentration $\hat{c}(t).$ At the upper limit, when only As atoms are supplying the droplet, the droplet concentration $\hat{c}(t)$ increases so that the NW length increases and the droplet decreases its volume. Similarly, at the lower limit, when due to solidification the liquid volume cannot fill a droplet with radius $r(t)$ and wetting angle $\beta(t+\Delta t)>\beta_{min},$ the solid phase will decrease the NW radius so as to obtain the unique $r(t+\Delta t)$ able to sustain the remaining volume at a wetting angle $\beta(t+\Delta t)=\beta_{min}.$
\end{enumerate}
\end{enumerate}

 As proposed above, the model has 3 parameters: the two diffusion lengths $\lambda_{\textrm{facet}}$ and $\lambda_{\textrm{SiO}_2}$ and the retro-diffusion factor $\epsilon$. Previous models consider $c^\star\simeq 0.01,$\cite{Glas2013} $\lambda_{\textrm{facet}}\simeq 1-5\mu m$\cite{Dubrovskii2006,Schroth2017} and $\lambda_{\textrm{SiO}_2}\simeq 50-90\ nm$\cite{Krogstrup2013,Shibata1999}.

\begin{figure}[h!]
  \includegraphics[width=15cm]{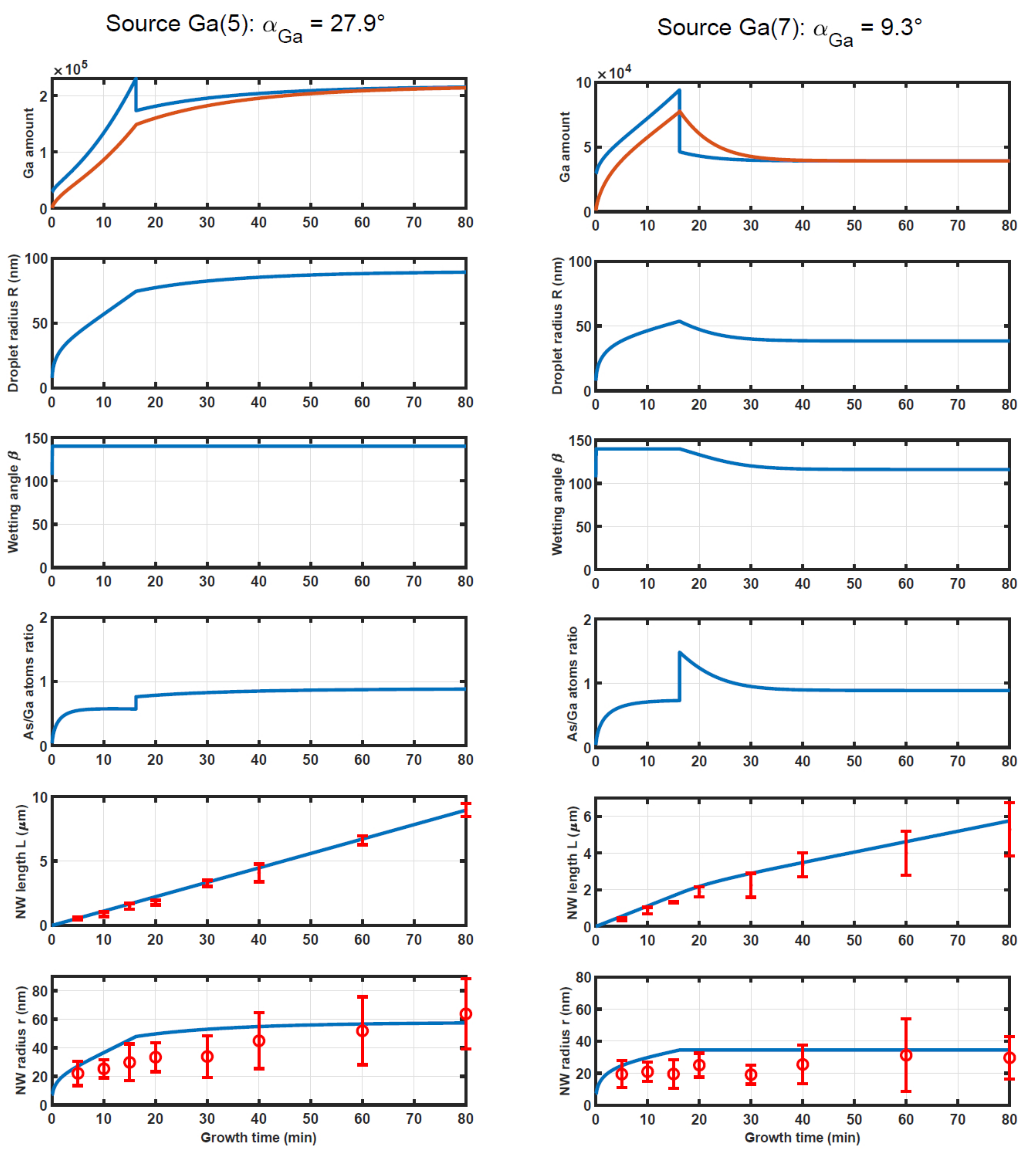}
  \caption{Numerical results (blues lines) obtained with the semi-empirical model for Ga(5) and Ga(7) sources. On the first line: the amount of Ga atoms supplying the droplet (in blue) and the amount of Ga atoms from the liquid droplet used for the NW growth (red) as function of the growth time. On the next lines: droplet radius, wetting angle, the As/Ga ratio supplying the droplet, NW length (computed-blue; experimental-red) and the NW radius (computed-blue and experimental-red). Numerical parameters are identified by fitting only the Ga(5) experimental data (left column). Using the same model parameters, but for the Ga(7) source, we have obtained the numerical results (blue lines) in the right column, plotted together with the experimental data (red points and error bars).}
  \label{fgr:Figure6}
\end{figure}

We have implemented the above described model with initial conditions $r(0)=7.5$ nm, $c(0)=c^\star=0.01$ and $\beta(0)=\pi/2,$ $L(0)=0$ and compute the evolution of the NW length, the NW diameter, the droplet size and the wetting angle as well as the amount of Ga and As atoms incorporated in the droplet during the process for the Ga(5) source. The best results, represented in Figure \ref{fgr:Figure6} (left) were obtained using the following parameters: $\lambda_{\textrm{SiO}_2}\simeq 70$ nm, $\lambda_{\textrm{facet}}\simeq 1.8$ $\mu$m and $\epsilon=0.127,$ in good agreement with previous cited references\cite{Dubrovskii2006,Schroth2017,Krogstrup2013,Shibata1999}. Parameters obtained by the best fit using the Ga(5)As NWs experimental data were subsequently used to predict the length and diameter evolutions of the Ga(7)As NWs. The results are reported in Figure \ref{fgr:Figure6} (right).

We are now able to explain the main differences induced by the source position: at very short times ($<1$ min), starting with identical NW geometry and droplet size as well as identical Ga fluxes on planar surfaces, since the amount of As atoms captured by the droplet is very small, both Ga(5) and Ga(7) droplets increase their volumes. Meanwhile, even in this regime, the As amount is sufficient to supply the axial growth of the NW. But since the amount of Ga atoms exceeds the amount of As atoms, the critical wetting angle is rapidly attained (at $\simeq t\simeq 20$ sec as shown in Figure 3 in SI) for both sources as the droplet radius $R(t)$ increases. As the NW length increases, the amount of Ga atoms supplying the droplet from diffusion over the NW facets from the Ga(5) source is about 3 times more important than that one from the Ga(7) source. This quantity becomes dominant for the droplet supplied by the Ga(5) source while for that one supplied by the Ga(7) source it has the same order of magnitude as the amount coming from diffusion on the substrate.

As shown in Figure \ref{fgr:Figure6}, due to the high $F_{As}/F_{Ga}$ ratio, the amount of As atoms supplying the droplet has the same order of magnitude than the amount of Ga atoms for both sources all along the growth process. This means that all As atoms supplying the droplet are transferred to the solid phase at each time step. But the remaining liquid phase contains less Ga atoms with the Ga(7) source than that with the Ga(5) source so, as a consequence, the increase of the NW diameter (under the droplet) of the Ga(7)As NWs is slower than that one of Ga(5) NWs. In turn, this implies that the size of the droplet for the Ga(5)As NWs increases faster than that one of the Ga(7)As NWs. The higher the droplet radius, the higher the amount of As atoms supplying the droplet, and this explains the faster axial growth of the Ga(5)As NWs with respect to Ga(7)As NWs for $t > 17$ min.

At $t\simeq 17$ min, corresponding to $L=1.8\ \mu$m, when the length of the NWs overcomes the diffusion length on the NW facets, a large amount of Ga supplying the droplet is gradually lost, but since the Ga flux on the NW facets with the Ga(5) source is higher than the Ga atoms lost for the NW growth, this is not a significative event for Ga(5)As NWs. For the Ga(7)As NWs, the As/Ga ratio becomes suddenly greater than 1 and, as a consequence, additional Ga atoms from the droplet will be used for solidification at each time step. As shown in Figure \ref{fgr:Figure6} (right), for Ga(7)As NWs the NW diameter stops to increase, the wetting angle decreases and the axial growth rate decreases accordingly.

The sudden lost of the Ga atoms supplying the droplet from substrate diffusion at $t\simeq 17$ min is actually a smoother transition between a regime dominated by the Ga atoms supplying the droplet from diffusion on the substrate and a regime dominated by the Ga atoms supplying the droplet from diffusion on the NW facets. Including this transition in the model will affect the local (in time) length and radius values but will have non-significant impact on the qualitative results.

These considerations highlight the importance of the Ga adatoms diffusion on the substrate, without which a large part of the Ga collected by the droplet would be missing and the experimental data could not be explained. Such a result is consistent with models previously developed by others\cite{Krogstrup2010,Krogstrup2013,Dubrovskii2005,Dubrovskii2006,Tchernycheva2007,Dubrovskii2008} but it should be considered as specific for the diffusion of Ga adatoms on $\textrm{SiO}_2$-terminated Si substrates, with a thin $\textrm{SiO}_2$ surface layer $1-2$ nm-thick, where the Ga adatom diffusion length is longer, whereas Ga adatoms can behave differently on thicker $\textrm{SiO}_2$ masks (typically $10-20$ nm-thick) used for substrate patterning\cite{Krogstrup2013}, as shown elsewhere\cite{Oehler2018,Vettori2018}.

It is interesting to notice that, in agreement with results in references\cite{Dubrovskii2008,Dubrovskii2015}, both classes of NWs evolve toward a {\em stationnary growth regime} when the amount of As and Ga atoms are identical and the growth mode is only axial. This asymptotic behavior is determined by two main factors: the fact that the V/III flux ratio is greater than 1 and the existence of a diffusion length for Ga adatoms along the NW facets. This is easily understood in a simplified framework when the NW radius is assumed constant but can be extended straightforwardly to variable NW radius growth models. Indeed, if the growth process is in a Ga-excess range, the droplet radius increases but since the V/III flux ratio is greater than 1, the system evolves toward a regime when the droplet is supplied with equal amounts of Ga and As atoms. At the opposite, in the As-excess range, the droplet decrease its volume and the direct flux amount on the droplet decreases for both species. However, since the droplet has an additional source of Ga atoms from NW facet diffusion, the system will evolve again toward a regime where the droplet is supplied with equal amounts of Ga and As atoms. These two arguments hold also when the NW radius evolves during the growth. The main reason for this is that the amount of atoms supplying the droplet from the direct flux scales (up to a bounded factor) like $r^2,$ while that one of atoms attaining the droplet through the diffusion on the NW facets scales like $r.$

\section*{Conclusions}

In conclusion, we experimentally demonstrated the influence of the incidence angle of the Ga flux on the growth kinetics of self-assisted GaAs NWs grown on Si$\textrm{O}_2$-terminated Si substrates. The experimental results demonstrate that this growth parameter significantly affects the NW length and diameter evolution. Subsequently, we develop a model and performed numerical simulation so as to fully explain the experimental results.

We developed a semi-empirical model and numerical simulations which highlight that the impact of the incidence angle of the Ga flux on the NW growth kinetics can be explained only by accounting for the contribution of Ga adatoms diffusing from the substrate surface to the Ga droplet. Such a result should be considered as specific for the diffusion of Ga adatoms on epi-ready Si$\textrm{O}_2$-terminated Si substrate, whereas Ga adatoms behave differently on patterned Si substrates with a thick Si$\textrm{O}_2$ mask.

The second equally important factor is the diffusion length of the Ga adatoms on the NW facets. The role of such a contribution to supply the Ga droplet becomes important when the NW length overcomes such a value, so that the droplet cannot be supplied anymore by the adatoms diffusing from the substrate. It then becomes the main contribution to the droplet supplying and, as expected, is depending on the Ga flux incidence angle. As a consequence, the difference in length and diameter between GaAs NWs grown with different Ga flux incidence angles can be explained assuming that variations in Ga supply may cause a different response from the Ga droplet between the two cases once the NW length exceeds the diffusion length of Ga adatoms on the NW facets. This will modify the volume and shape of the Ga droplet, thus affecting the surface of capture of As atoms and consequently the NW axial growth rate and also the NW diameter when the wetting angle of the Ga droplet becomes equal to a maximum value of typically $140^\circ$.

Ultimately, the results here reported show that the incidence angle of the Ga flux is an essential parameter to obtain good control over the self-assisted GaAs NWs grown by VLS-MBE. Such a result is quite significant, since it opens up to the possibility, having Ga cells with properly different incidence angles, of achieving fine control over the NW geometry and probably also over the NW crystal structure, by quickly modifying the amount of the incident Ga flux and therefore the amount of Ga supplying the droplet.

\section*{Experimental section}

The samples subject to this study were realized in a MBE reactor Riber 32 equipped with two Ga cells with different flux incidence angles respectively equal to $27.9^\circ$ (denoted Ga(5) cell) and to $9.3^\circ$ (denoted Ga(7) cell), and an $\textrm{As}_{4}$ valved cracker cell with a flux incidence angle equal to $41^\circ$. All substrates employed for the growths consisted of $1\times 1$ $\textrm{cm}^2$ chips of boron-doped Si(111) (0.02-0.06 $\Omega\cdot$cm) with an epi-ready surface oxide layer ($\simeq 1-2$ nm-thick). The substrates were cleaned by sonication in acetone and ethanol for 10 min and degassed at $200^\circ$C in ultra-high vacuum before introduction in the MBE reactor. In all cases 1 ML of Ga was pre-deposited at $520^\circ$C always with the Ga(5) cell so as to form Ga droplets and, subsequently, pins into the surface oxide layer when the substrate temperature is increased\cite{Madsen2011,Tauchnitz2017}. The substrate temperature was subsequently increased up to $610^\circ$C in 10 min and stabilized for 2 min. Then the substrate was exposed to Ga and $\textrm{As}_4$ fluxes. As far as Ga is concerned, the flux in question was originated either by the Ga(5) or the Ga(7) cell, but in any case the Ga flux adopted corresponded to a planar growth rate equal to 0.5 ML/sec, defined in terms of equivalent growth rate of a 2D GaAs layer grown on a GaAs substrate, as measured by reflection high energy electron diffraction (RHEED) oscillations. Similarly, the $\textrm{As}_4$ flux was equal to an equivalent 2D GaAs layer growth rate\cite{Rudolph2011} of 1.15 ML/sec, thus providing an As/Ga flux ratio $F_{As}/F_{Ga}= 2.3$ for a GaAs growth on the substrate. The NW growths were finally stopped by closing the shutter of Ga and $\textrm{As}_4$ cells simultaneously and fast decreasing of the sample temperature, so as to preserve the Ga droplet on the NW top.

\begin{acknowledgement}
The authors want to acknowledge the French Agence Nationale de la Recherche (ANR) for its funding to the HETONAN project (http://inl.cnrs.fr/projects/hetonan/). The authors thank also J. B. Goure and C. Botella for technical assistance and the NanoLyon platform for access to the equipments.
\end{acknowledgement}

\begin{suppinfo}
%
%
The following files are available free of charge.
\begin{itemize}
  \item {The file VettoriSI.pdf} : Supporting Information file contains details for : the samples grown at different incidence angles of the Ga flux, Ga deposition and NW nucleation, images of NWs grown for short times and the algorithm for the implementation of the model.
\end{itemize}

\end{suppinfo}

\providecommand{\latin}[1]{#1}
\makeatletter
\providecommand{\doi}
  {\begingroup\let\do\@makeother\dospecials
  \catcode`\{=1 \catcode`\}=2 \doi@aux}
\providecommand{\doi@aux}[1]{\endgroup\texttt{#1}}
\makeatother
\providecommand*\mcitethebibliography{\thebibliography}
\csname @ifundefined\endcsname{endmcitethebibliography}
  {\let\endmcitethebibliography\endthebibliography}{}

\end{document}